\documentclass{optica-article}

\journal{opticajournal} 

\articletype{Research Article}

\usepackage{textcomp}
\usepackage[normalem]{ulem}
\usepackage{float}
\usepackage[percent]{overpic}
\usepackage{multirow}

\usepackage{lineno}

\begin{document}

\title{Development towards high-resolution kHz-speed rotation-free volumetric imaging}

\author{
Eleni Myrto Asimakopoulou,\authormark{1,*} 
Valerio Bellucci,\authormark{2,*} 
Sarlota Birnsteinova,\authormark{2}
Zisheng Yao,\authormark{1}
Yuhe Zhang,\authormark{1}
Ilia Petrov,\authormark{2}
Carsten Deiter,\authormark{2}
Andrea Mazzolari,\authormark{3,4}
Marco Romagnoni,\authormark{3,4}
Dusan Korytar,\authormark{5,6}
Zdenko Zaprazny,\authormark{6}
Zuzana Kuglerova,\authormark{7}
Libor Juha,\authormark{7}
Bratislav Luki\'c,\authormark{8}
Alexander Rack,\authormark{8}
Liubov Samoylova,\authormark{2}
Francisco Garcia Moreno,\authormark{10,11}
Stephen A. Hall,\authormark{12}
Tillmann Neu,\authormark{10,11}
Xiaoyu Liang,\authormark{13}
Patrik Vagovic,\authormark{2,9}
and Pablo Villanueva-Perez\authormark{1}}

\address{
\authormark{1}Synchrotron Radiation Research and NanoLund, Lund University, Sweden\\
\authormark{2}European XFEL GmbH – The European X-Ray Free-Electron Laser, Schenefeld, Germany\\
\authormark{3}Ferrara University, Department of Physics and Earth Science, Via Saragat 1, 44122 Ferrara, Italy\\
\authormark{4}INFN, Ferrara Division, Via Saragat 1, 44122 Ferrara, Italy\\
\authormark{5}Integra TDS, s.r.o., Krakovany, Slovakia\\
\authormark{6}Institute of Electrical Engineering SAS, Dúbravská cesta 9, 841 04 Bratislava\\
\authormark{7}Department of Radiation and Chemical Physics, Institute of Physics, Czech Academy of Sciences, Prague, Czech Republic\\
\authormark{8}ESRF - The European Synchrotron, Grenoble, France\\
\authormark{9}Center for Free-Electron Laser Science (CFEL), DESY, Hamburg, Germany\\
\authormark{10}Institut für Werkstoffwissenschaften und -technologien, Technische Universität Berlin, Hardenbergstr. 36 10623 Berlin\\
\authormark{11}Helmholtz-Zentrum Berlin für Materialien und Energie GmbH (HZB), Hahn-Meitner-Platz 1, 14109 Berlin\\
\authormark{12}Division of Solid Mechanics, Lund University, Ole
Römers Väg 1, 223 63 Lund, Sweden\\
\authormark{13}Institute of Multidisciplinary Research for Advanced Materials (IMRAM), Tohoku University, 2-1-1 Katahira, Aoba-ku, Sendai, Miyagi 980-8577, Japan\\
}

\email{\authormark{*}eleni\_myrto.asimakopoulou@sljus.lu.se, valerio.bellucci@xfel.eu} 


\begin{abstract*}

X-ray multi-projection imaging (XMPI) provides rotation-free 3D movies of optically opaque samples. 
The absence of rotation enables superior imaging speed and preserves fragile sample dynamics by avoiding the shear forces introduced by conventional rotary tomography.
Here, we present our XMPI observations at the ID19 beamline (ESRF, France) of 3D dynamics in melted aluminum with 1000 frames per second and 8~$\mu$m resolution per projection using the full dynamical range of our detectors.
Since XMPI is a method under development, we also provide different tests for the instrumentation of up to 3000 frames per second. 
As the flux of X-ray sources grows globally, XMPI is a promising technique for current and future X-ray imaging instruments.

\end{abstract*}


\section{Introduction}
Time-resolved X-ray imaging has been a fundamental technique for non-destructive exploration of phenomena happening in samples opaque to visible light. 
The speed and performance of 2D X-ray imaging (radiography) have constantly increased during the last century, with improving cameras and X-ray sources. 
State-of-the-art radiography can now record with MHz frame rate speeds at high-brilliance sources such as storage rings~\cite{APS:2008, Olbinado:2017} and X-ray free-electron lasers~\cite{Vagovic:2019}.
On the other hand, time-resolved 3D imaging, also known as tomoscopy, has only recently reached 1~kHz at storage rings\cite{Moreno:2021}. 
Indeed, the maximum acquisition speed for tomography is limited by the need to rotate the sample, so it induces shear forces that can alter the studied processes. 
For this reason, it is today still very difficult to capture 3D phenomena in samples opaque to visible light with under-millisecond time resolution. 
The investigation of many fast phenomena would benefit from this capability, including samples of scientific and industrial interest such as fractures in solids \cite{Kumar2016StrengthIO, XU2020104165}, shock wave propagation \cite{ShockWP:2016, Shockwave:1998}, fast biological processes \cite{Hansen2021, Truong2020} and fluid dynamic studies \cite{MultiPhaseFlowXRaysReview2021}. 
To overcome this problem, there have been developments in the field of X-ray Multi-Projection Imaging (XMPI), which allow for rotation-free 3D imaging \cite{Pablo:2018, Duarte:2019, Yashiro:2020, DLS:2021-2023, patent:Multi-projection}. 
XMPI divides the X-ray beam into multiple beamlets to enable imaging of a sample from various angles. 
The recorded sparse dataset of images is then fed into advanced reconstruction algorithms to obtain a full 3D reconstructed image\cite{ONIX:2023}. 
Thus, the XMPI method enables the study of dynamics without introducing shear forces that may alter the studied processes~\cite{centrifugal:2000, Moreno:2021}. 
In this paper, we demonstrate static and time-resolved XMPI with a frame rate of up to 3~kHz without focusing the beam of ID19 at the European Synchrotron Radiation Facility (ESRF). 
XMPI was then used for the study of fast stochastic processes in aluminum foam samples. Such samples have been used before for 1~kHz frame-rate tomographic acquisition at synchrotron facilities \cite{Moreno:2021}. 
Several of the processes of interest that take place in the foam are highly sensitive to the g-forces generated by the fast rotations needed to probe the relevant dynamics, which motivates the use of XMPI for these studies over tomography at higher frame rates. 
Recently, there have been pioneering attempts to employ XMPI with white beam using different splitting configurations at storage rings~\cite{Yashiro:2023,voegeli2023}.
Furthermore, a 1.1 MHz sampling method of XMPI was also demonstrated with a two-beamlet configuration~\cite{MHzArxiv2023}.
Here we present the first use of XMPI with a pink beam for application in diffraction-limited storage rings. 
The narrow energy bandwidth results in images carrying quantitative information on the sample, as well as sharper images compared to a white beam produced by a bending magnet or wiggler.

\begin{figure}[htbp]
\centering
\includegraphics[width=1\linewidth]{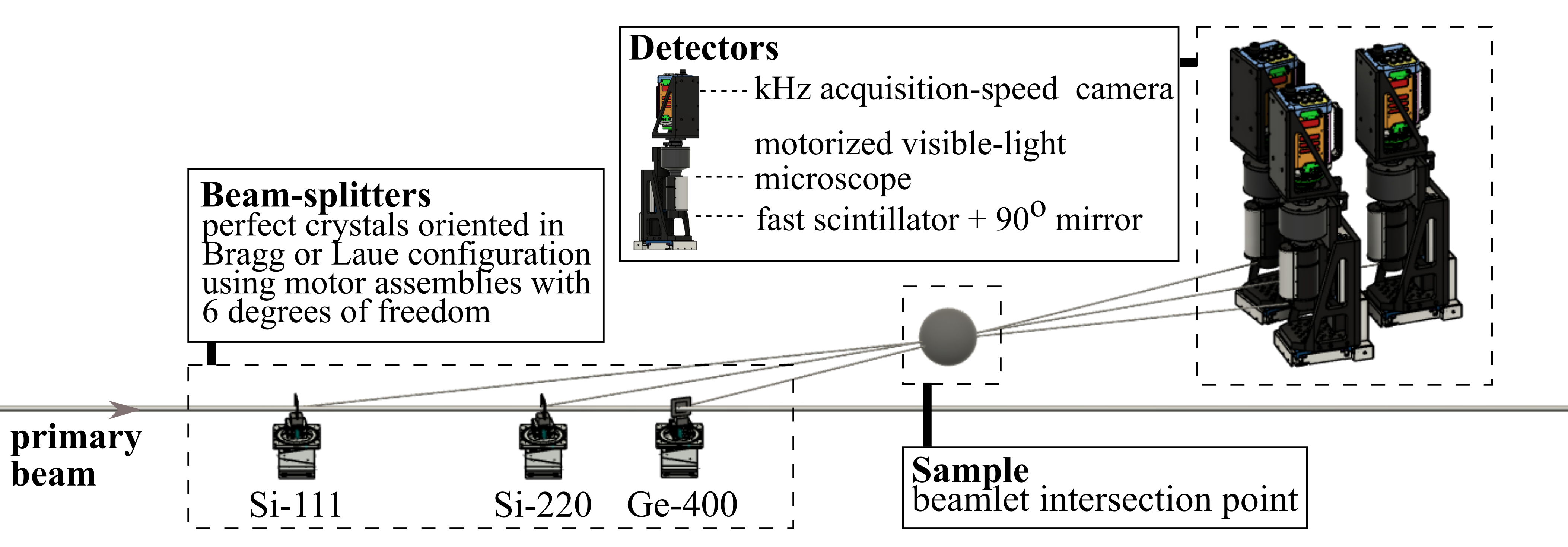}
\caption{
Illustration of the XMPI concept \cite{patent:Multi-projection}, as implemented at ID-19 of ESRF. The main X-ray beam enters the experimental hall from the left-hand side, propagating to the right of the figure. Three beam splitters are oriented in Bragg or Laue configuration (details in the main text) to meet the Bragg diffraction condition and deflect a part of the main beam. The beam splitters are positioned so that the deflected beams intersect at one point, where the sample is placed. The detectors are placed behind the sample, each aligned to one of the deflected beams, to image the sample simultaneously from different directions.
}  
\label{fig:schematic}
\end{figure}

\section{Methods}
The XMPI method relies on crystal diffraction for the splitting of a primary X-ray beam into secondary beams in order to enable simultaneous imaging of a sample from angularly spaced viewpoints. 
The configuration of the XMPI setup is flexible and can be optimized to different facilities.
The ID19 ESRF beamline was chosen for this demonstration of XMPI because its undulator source in a diffraction-limited storage ring and grants one of the highest spectral densities available for synchrotron sources, 2-3 orders of magnitude higher than beamlines based on bending magnets or wiggler sources \cite{ID19:2010}.
In general, access to a high-brilliance source is crucial for fast imaging due to the small exposure time for each frame. 
In the case of XMPI, high spectral density is important because the beam is divided spectrally to produce multiple projections. 
The concepts of the presented implementation of XMPI are introduced in~\ref{subsec:xmpiSetup}. The required methods for the optimization of the setup at the conditions of ID19 are motivated and described in~\ref{subsec:setupAlignment},~\ref{subsec:crystals} and~\ref{subsec:thermalCamera}.

\subsection{XMPI setup}
\label{subsec:xmpiSetup}
The experimental setup is composed of three main parts; the multi-projection generation, the sample environment, and the detector system.
A schematic view of the setup is depicted in Fig.~\ref{fig:schematic}.
The multi-projection generation is achieved with the use of crystals, which are referred to as ``beam splitters''. 
Each beam splitter is mounted on motorized stage assemblies with six degrees of freedom (three linear translations and three angles), which allows the alignment of the setup. 
During alignment, each crystal is positioned in the beam path and oriented at an angle $\theta$ with respect to the central beam to fulfill its respective Bragg or Laue condition for a specific part of the spectrum. 
When the central beam interacts with an aligned beam splitter, part of the beam is diffracted, creating a secondary beam that travels at a $2\theta$ angle with respect to the central beam and is referred to as a ``beamlet''. 
The portion of the beam that is neither diffracted nor absorbed by a given beam splitter is transmitted. 
The arrangement of the beam splitters is such that the beamlets intersect at a common point, offering an equal number of projections on the sample. 
The setup for this work relied on three beam splitters (yet this number will increase in future iterations of the experiment). 
The exact configuration was finalized after the beam characterization (see Sec.~\ref{subsec:setupAlignment}), as it is further discussed in Sec.~\ref{subsec:crystals}. 
The sample environment is placed at the beamlet intersection point, allowing its simultaneous imaging from the three generated viewpoints. 
Because of its design, the XMPI method can accommodate a wide variety of sample environments.
The imaging projections are detected with three indirect detectors, each aligned perpendicular to a beamlet's projection direction. 
All indirect detectors were identical and were composed of (i) an efficient scintillator ($250~\mu$m thick GAGG+), (ii) a $90^{\circ}$ mirror which redirects the visible light to a motorized microscope configuration, (iii) a magnification system consisting of a visible microscope, which also focuses the light on the camera sensor, and (iv) fast cameras optimized for kHz acquisition speed (Photron Nova S16 models). 
Each microscope used the same configuration of optics, a 5X Mitutoyo Plan Apo HR Infinity Corrected Objective, and a 1x MT-L tube lens. 
The described configuration provided a field-of-view of $4\times 4$~mm$^2$, with an effective pixel size of 4~$\mu$m. 


\subsection{Setup alignment}
\label{subsec:setupAlignment}
XMPI is based on crystal diffraction and its configuration is inherently linked to the spectrum and the central energy of the beamline, as it determines the arrangement of the setup.  
The central energy of ID19 U17.6 undulator at the minimum gap of $11.8$~mm ($\lambda_{u}=18$~mm, N=90) was expected to be approximately $17.6$~keV on-axis with 1.6\% bandwidth (FWHM). 
These values were obtained from beamline simulations with XOP utilizing parameters of the EBS lattice \cite{XOP:2011, Raimondi:2023}.  
The beam was filtered only with mandatory optical elements for heat load moderation (0.8~mm thick diamond window and a series of thin Be windows $\approx$2~mm), cutting out the soft part of the emitted X-ray spectrum ($<$ 10~keV) along the 145~mm vacuum flight tube.  
We implemented a spectrometer to measure the exact beam spectrum and spectrally align the crystals accordingly, i.e. set the crystals in diffraction conditions to the high flux components of the spectrum. 
The spectrometer was placed as the most downstream element of the setup to act as a monitoring tool~\cite{Ilia:2021}.
The spectrometer was composed of a bent C-$333$ crystal, an energy-dispersive element that diffracts different energies with different angles, coupled to a 1x indirect detector PCO.edge.5.5 camera. 
The diffracted photons are therefore detected in different positions (pixels) in the camera depending on their energy, which allows the monitoring of the beam spectrum. 
The measurement of the beam's central energy was carried out by first aligning the spectrometer to diffract the most intense part of the beam spectrum and then using a perfect crystal to gauge the angle where its diffraction condition was fulfilled. 
A perfect silicon crystal with a <100> out-of-plane direction was used for this purpose and was positioned upstream of the spectrometer. 
The crystal was rocked continuously in one direction until it was oriented to satisfy the Laue diffraction from the (220) diffraction plane. The central beam spectrum was simultaneously being monitored in the PCO.edge.5.5 camera.
While the silicon crystal was far from satisfying the diffraction condition, the beam spectrum remained unchanged (aside from uniform intensity losses due to transmission through the silicon crystal). 
When the crystal was rocked in the diffraction region, a clear dip was visible in the camera, indicating the portion of the beam that was being diffracted by the silicon crystal. 
The dip moved through the beam spectrum as the crystal was rocked through the diffraction region, indicating that the crystal was satisfying the diffraction condition for a different part of the spectrum. 
The same procedure was repeated with the rocking of the crystal in the opposite direction. 
The two rocking angles that satisfied the diffraction condition at the most intense part of the spectrum for each of the rocking directions need to be symmetric and were therefore used to assess accurately the diffraction angle for the crystal in the beam's central energy. 
With this approach, the central energy of ID19 was estimated to be $17.25$~keV with the smallest undulator gap, i.e., the highest flux available with 7/8 filling mode and 200~mA storage ring current.

\subsection{Beam splitters configuration}
\label{subsec:crystals}
The efficiency of the setup is directly affected by the choice of the crystals that act as the beam splitters and beam parameters (intensity, divergence, spectral components). 
We positioned different crystals one by one upstream of the calibrated spectrometer in order to characterize their spectral response, i.e., how much of the ID19 beam they diffract.  
Silicon and Germanium monocrystals were assessed to be optimal for the pink beam of ID19.
The final selection of beam splitters is summarized in Tab.~\ref{tab:crystals}, which will hereafter be referred to using their abbreviations. The crystals were provided by INFN-Ferrara.

\begin{table}
    \centering
    \begin{tabular}{c|c|c|c}
    & Beam splitter \#1 & Beam splitter \#2 & Beam splitter \#3 \\ 
    \hline    
    Material & Silicon & Silicon & Germanium \\
    Thickness & $14$ µm & $14$ µm & $1$ mm \\
    Out-of-plane & <211> & <100> & <100> \\
    Diffraction plane & (111) & (220) & (400) \\
    Diffraction & Laue & Laue & Bragg \\
    Diffraction angle & 6.58$^\circ$ & 10.79$^\circ$ & 14.72$^\circ$ \\
    Projection angle & 13.16$^\circ$ & 21.58$^\circ$ & 29.44$^\circ$ \\
    Abbreviation & Si-111 & Si-220 & Ge-400 
    \end{tabular}
    \caption{Configuration of the beam splitters for XMPI at beam energy $17.25$~keV. The splitters' enumeration (\#1-\#3) is the order of interaction with the beam. All the beam splitters are mono-crystals. The listed abbreviations are used to refer to the beam splitters in the main text.}
    \label{tab:crystals}
\end{table}

\subsection{Heat–load measurements}
\label{subsec:thermalCamera}
\begin{figure}[htbp]
\centering
\includegraphics[width=0.7\textwidth]{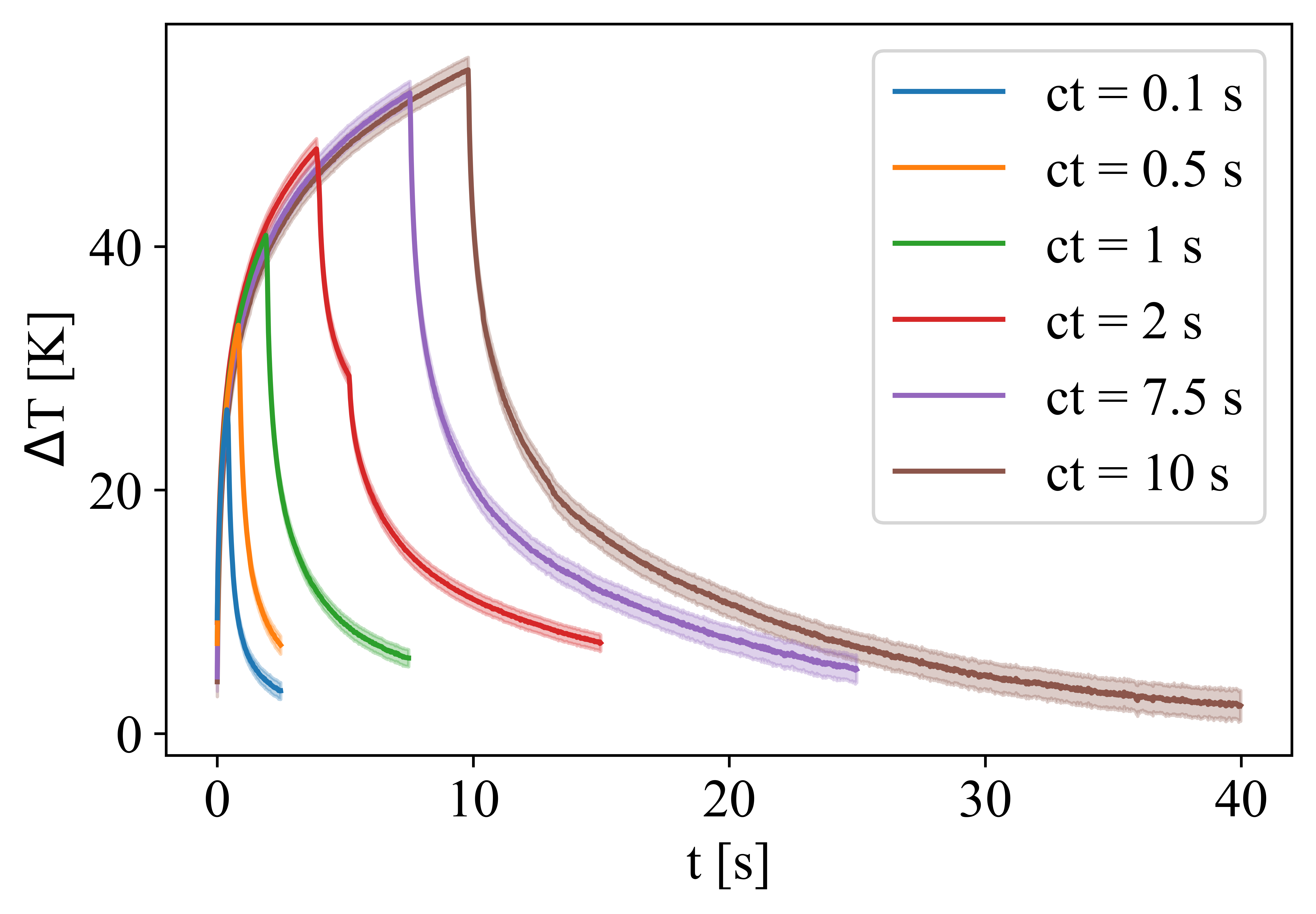}
\caption{
Measured temperature change $\Delta{T}$ versus time $t$ on the last crystal Ge-400 for different times of opened shutter ct. The error bar is visualized as a shade. The temperature change is far from the melting point of germanium T = 1209 K. 
}  
\label{fig:thermo}
\end{figure}

The final configuration of beam splitters in the experiment included the use of a germanium crystal (see Tab.~\ref{tab:crystals}). Ge-400 crystal was oriented in Bragg and absorbed the entire beam that was not diffracted or absorbed by the previous elements (see Fig.~\ref{fig:schematic}).
Given the high flux absorbed by this crystal, there were concerns about the thermal dissipation of the absorbed heat load within the crystal and we monitored its surface temperature via a thermal camera (FLIR A7600sc). 
Fig. \ref{fig:thermo} shows the temporal evolution of the temperature registered by the FLIR camera at different exposure times controlled by a shutter (ct) for various time periods. 
The FLIR camera provides a thermal image as well, which reveals a spatial distribution of temperatures on the irradiated and surrounding surfaces. 
The temperature rise can be estimated from measured relative intensity $\Delta{C}$ of near-infrared thermal radiation using the proper value of relative emissivity of germanium $\varepsilon$. 
Emissivity is temperature dependent and thus may vary during the crystal heating. 
The value of relative emissivity $\varepsilon$ for room temperature Ge was measured as $\varepsilon$ = 0.26 in \cite{Madura:1997} and for T = 1209 K, i.e. melting point of Ge, as $\varepsilon$ = 0.55 in \cite{Allen:2004}. 
Since the experiment was performed at room temperature and the change of measured relative intensity was around 2 \% of the whole dynamic range, we conclude that the value of relative emissivity coefficient $\varepsilon$ = 0.26 reported by \cite{Madura:1997} for room temperature fits well the conditions of the present experiment. 
The measured increase in the temperature of the Ge crystal (Fig. \ref{fig:thermo}) is of the order of tens of degrees and well below the melting point of Ge. 
No distortion is visible in the images from the Ge-400 projection. 
This includes distortions due to thermal strain caused by overheating the crystal. 
Therefore, we can conclude that the heat transfer from the crystal surface to the crystal bulk, the air, and the crystal holder is effective enough to prevent the overheating of the irradiated region.

\section{Results and Discussion}

\begin{figure*}[htbp]
\centering
\includegraphics[width=1\textwidth]{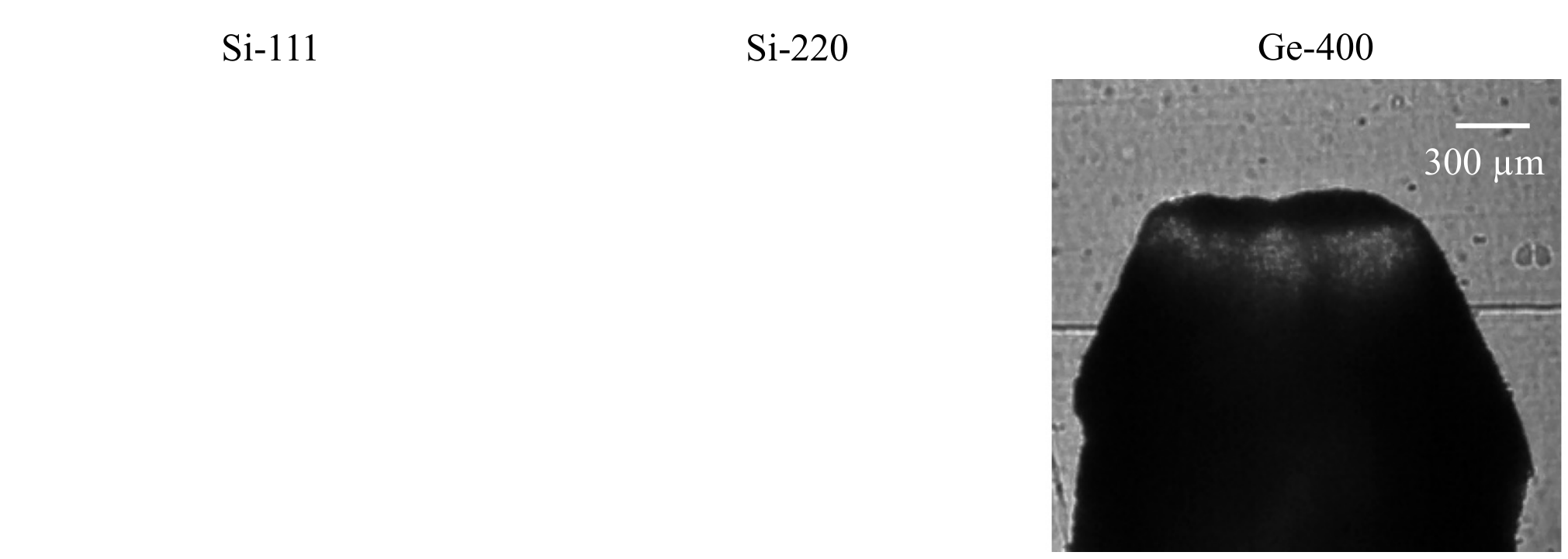}
\caption{
Simultaneous frame acquisition of an alignment pin at $3000$~fps speed from three projections, coming respectively from the beam splitters Si-111, Si-220, Ge-400. 
The Si-111 and Si-220 beam splitters are in Laue geometry, while the Ge-400 is in Bragg. 
The blurring in the Si-111 and Si-220 projections is caused by the formation of the Borrmann triangle combined with some warping in these thin membrane-like splitters.
}
\label{fig:3kHz_static}
\end{figure*}

An alignment sample (bamboo toothpick) was used to determine and monitor the intersection of the Si-111, Si-220, and Ge-400 beamlets. 
The final configuration allowed recording up to 3 kHz with the three cameras simultaneously using the whole 12-bit dynamical range of the kHz cameras. 
This corresponds to frame acquisitions every $0.3$ ms, allowing the reconstruction of sub-millisecond dynamics. 
The microscope configuration has 8~$\mu$m spatial resolution per frame, estimated using a resolution target. 
The recorded images with all three cameras can be seen in Fig.~\ref{fig:3kHz_static}. The Si-111 image in Fig.~\ref{fig:3kHz_static} was rotated by $5.81^\circ$ to recover a misalignment error in the camera.

The illumination of the alignment sample allowed some observations about the employed beam splitters.
The Ge-400 projection appears intense and clear and is used as a reference for the capabilities of the setup. 
However, the Si-111 and Si-220 projection images appear blurry (Fig.~\ref{fig:3kHz_static}).
After dedicated testing, it was concluded that the cause of the blur in the Si-111 and Si-220 was the result of two factors: the Laue diffraction geometry and the presence of strain in the crystal. 
The Laue diffraction geometry incurs into blurring caused by the Borrmann triangle \cite{Authier:book}, a dynamical diffraction effect. 
Indeed, an X-ray beam diffracted over the entire thickness of a crystal results in a distribution of the diffraction centers over a fan-shaped profile, which produces a blurry virtual source. 
Laue diffraction is especially sensitive to strain in the crystal since strained areas are illuminated differently under the profile of the fanned diffracted beam. 
An example of this can be seen at the top of the Si-220 projection image, where a halo appears. 
The measurements were carried out with a low-strain area of the crystal to mitigate these effects. 
The blurring induced by the Borrmann triangle does not happen for symmetric Bragg diffraction geometry, as visible in the Ge-400 image, since the thickness of crystal traversed by the X-rays before diffraction is extremely reduced. 
Therefore, we advise using symmetric Bragg diffraction geometry when possible for future iterations of XMPI experiments. 

In the Si-220 projection, the significant strain produces a curvature in the crystal shape that results in warping of the beam trajectories, stretching the recorded projections along this beam. 
Because of this stretching and the aforementioned Borrman-triangle effect, the Si-220 figure was re-scaled over the horizontal direction by a factor of 1.17 to match the real aspect ratio of the object. 
The strain was most likely introduced during fabrication or by the clamping method. 
Both causes are technological in nature and they can be improved upon with further engineering of the crystals and their holders.
Regarding the Si-111 splitter, we observe strong vibrations as a result of the thickness and mounting of this membrane. 
In future iterations, we will redesign the mounting and use a thicker crystal to avoid such vibrations.

To demonstrate the capabilities of XMPI at ID19, we explore the foaming of an aluminum alloy produced by a fast and self-triggering pressure release method. 
The complete foaming and solidification process of the alloy takes place in less than one second and has never been studied before at such speeds. 
Rapid metal foaming and solidification not only have the advantage of short production time but also of rapid development, as the foam structure has much less time to coarsen. 
Time is one of the most important parameters, along with temperature and alloy composition, leading to an increased number of coalescence events \cite{GARCIAMORENO2020325}. 
Long foaming times lead to an increased number of coalescence events and, thus, to foam aging.  

This process demonstrated dynamics at the order of 0.5-1 kHz, so we optimized our acquisition system to that range. 
Although similar samples have been observed in 3D at kHz frame rate during a standard metal foaming process, it was also observed that in some cases, the dynamics of low-stabilized liquid metal foams such as the alloy studied in this work were altered in the later foaming stage by the radial forces generated by the rotation \cite{Moreno:2021}. 
Thus, XMPI is a promising technique for their study without sample rotation and, therefore, without altering the dynamics of interest. 
The studied sample was produced from a foamable thixocast precursor made of an AlSi6Cu4 alloy with an added 0.8 wt\% TiH2 as a blowing agent. 
Thixocast foamable precursors are known for their high volume expansion and capability to form round bubbles but also for their tendency for bubble coalescence \cite{Weise_Stanzick_Banhart_2003}. 
By observing the evolution of such liquid metal foams in 3D, we gain insight into fast phenomena during the foaming process, such as the evolution of film thickness, i.e. the distance between bubbles, over time, the minimum film thickness, gas interdiffusion between different bubbles, or the rupture and coalescence of bubbles orders of magnitude faster than normal foam expansion.

The measurements were carried out with two out of the three beam splitters (Si-220 and Ge-400) of the final XMPI configuration since the illumination from the Si-111 beam splitter was experiencing intense drifting, making the recording unusable. 
During the measurements, the expansion of the foams occurs rapidly due to the fast intrinsic pressure release, making it non-trivial to determine the correct time to trigger the acquisition.
The detector initially assigned for the Si-111 crystal was repurposed to provide a visual trigger to begin the acquisition process. 
A laser was set to illuminate the sample so that the visible light image captured on the camera had the same field-of-view as the X-ray images provided by the other two cameras, i.e. it was a close representation of the expected X-ray images. 
Therefore, the view from this camera was used to manually trigger the acquisition of the two other cameras when the expanding foaming was visible in the field of view. 

\begin{figure*}[htbp]
\includegraphics[width=1\textwidth]{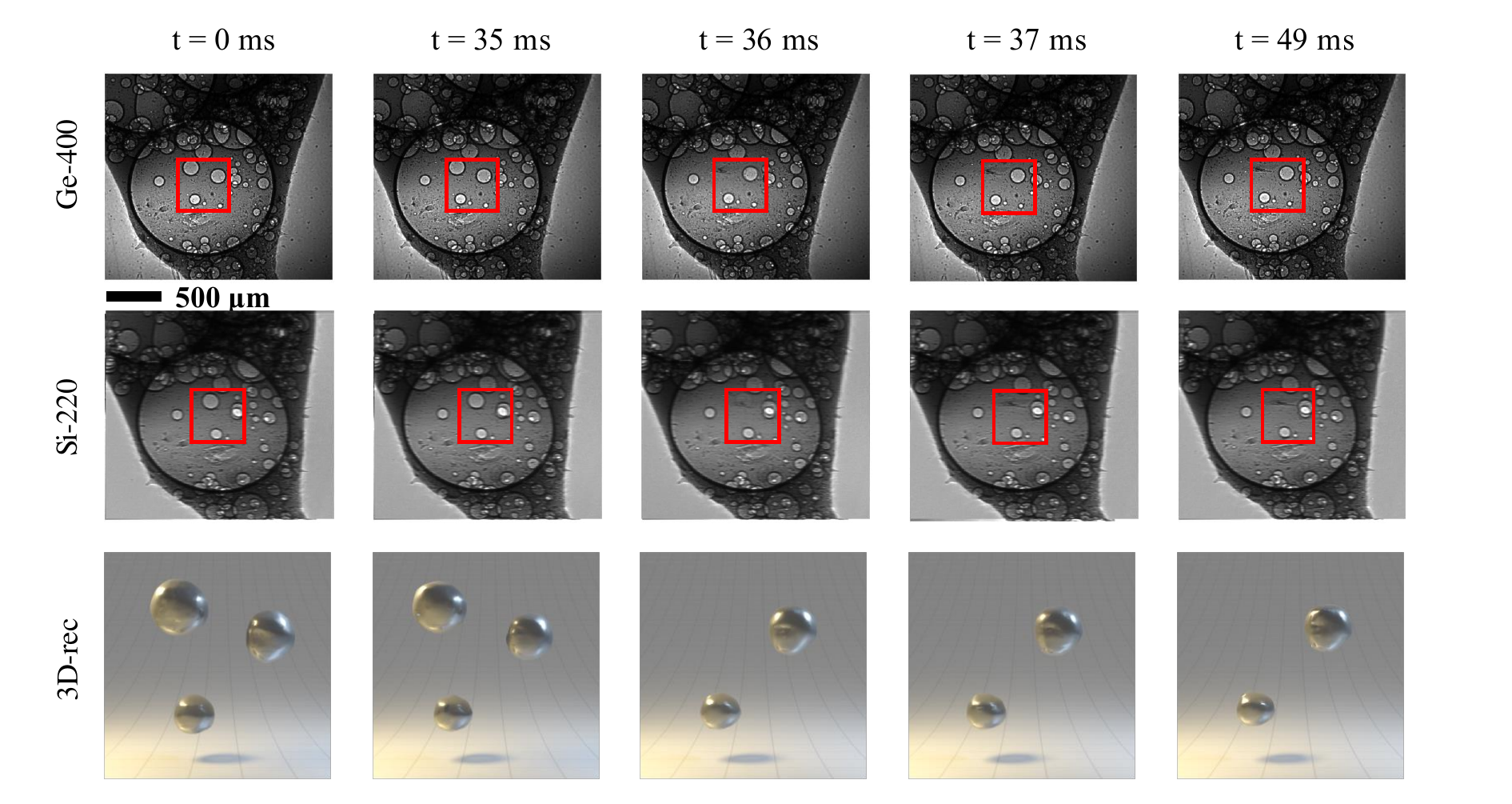}
\caption{
A selection of frames was collected with two projections of the XMPI method during the foaming of an AlSi6Cu4 thixocast precursor. 
The red square marks the region of interest for the reconstruction, while the first (T=$0.000$s) and last (T=$0.049$s) timestamps mark the time period that was used for the reconstruction. 
In the interim consecutive frames (T=$0.035$s, T=$0.036$s and T=$0.037$s), behaviors in the bubble movement can be monitored. 
The last row shows renderings of 3D reconstructions of the selected bubbles. 
At t = 36 ms, the lower left bubble burst with no significant effect on the rest of the bubbles' position, volume, or rearrangements in the neighborhood. 
 }  
\label{fig:frames}
\end{figure*}

The 3D structure of metallic foams was reconstructed using a deep learning 3D reconstruction approach. 
We applied ONIX~\cite{ONIX:2023}, a self-supervised learning method developed to address the ill-defined problem posed by XMPI. 
A continuous function~\cite{Mildenhall2020NERF,yu2021pixelnerf} was learned that maps from the spatial-temporal coordinates to the refractive index of the sample, depicting the structure of the metallic foams.
Given the limited volumetric data available from the two projections, we adopted adversarial learning~\cite{goodfellow2020generative} to enforce the self-consistency between the collected projections and the predictions generated from arbitrary viewpoints. 
An additional discriminator as the one used in \cite{ledig2017photo} was added to the ONIX architecture, and the networks were trained using the adversarial loss function\cite{goodfellow2020generative,schwarz2020graf}. 
We selected a region of interest with the size of 128$\times$128 on both projections for the 3D reconstruction, as shown in Fig.~\ref{fig:frames}. 
The selected areas were denoised using total variation filters and segmented from the background before feeding into the neural networks. 
The network training was performed on an NVIDIA A100 GPU with 40 GB of RAM. 
We used Adam optimizer~\cite{kingma2014adam} with a learning rate of 10$^{-4}$ and batch size of eight.
After the training, the volumetric information was obtained by extracting 128$^3$ grids from the networks, with the voxel size equivalent to the effective pixel size of 4~$\mu$m. 
The resulting 3D reconstructions were visualized using Blender.
A selection of frames from the two projections can be seen in Fig.~\ref{fig:frames}. 
In the shown projections, slight bubble movements are visible and the rupture of a surface bubble can be observed, which results in a dark corrugated structure at its former position and will be discussed later in more detail. 
More accurate results are visible in the reconstructed 3D volume. 

\begin{figure*}[htbp]
\includegraphics[width=1\textwidth]{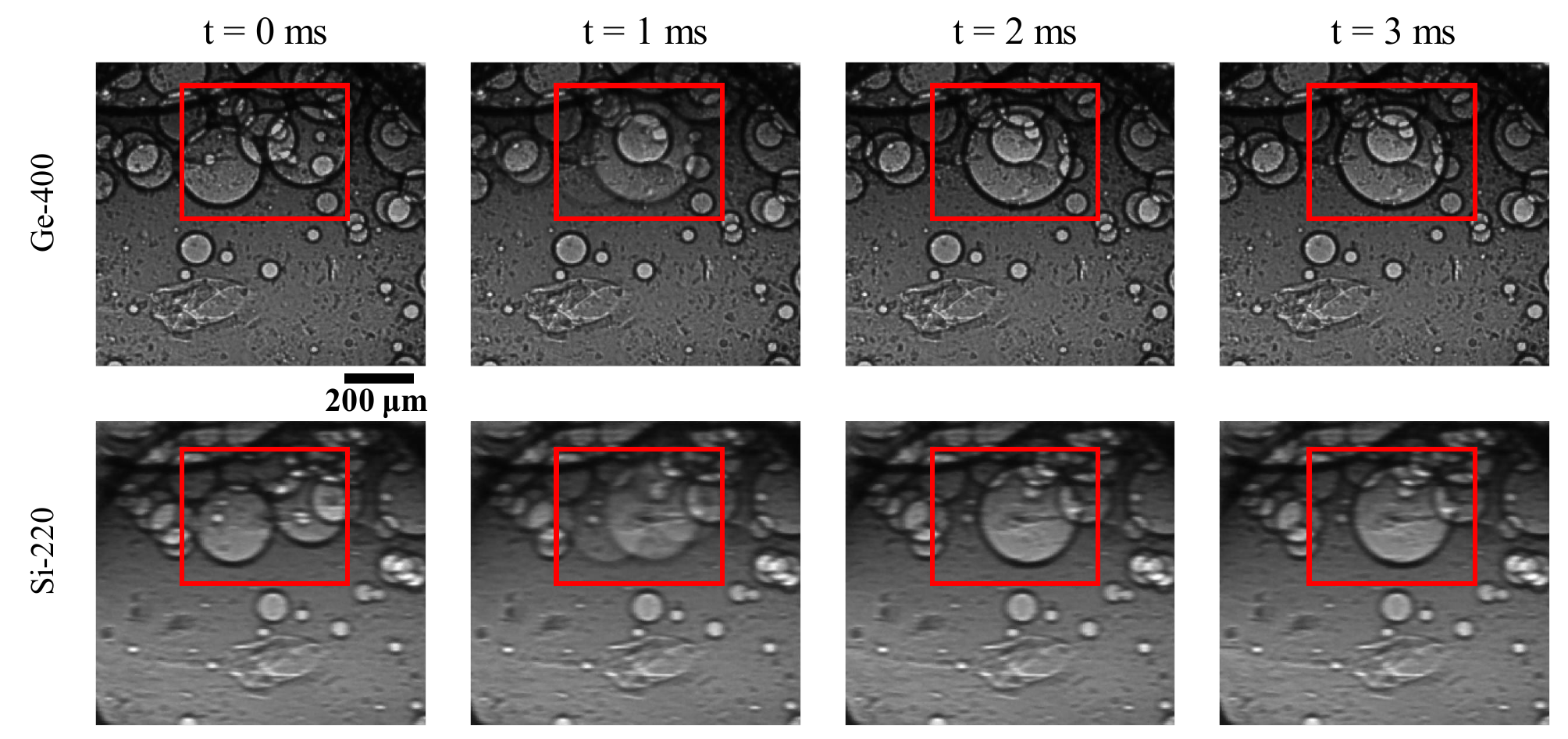}
\caption{
Bubbles merging in the melted aluminum foam sample. 
The selection of frames shows two similarly sized bubbles merging into one larger one, details marked by the red square. 
This process takes a maximum of 1-2 frames, i.e. of 1-2 ms. 
 }  
\label{fig:merging}
\end{figure*}

When the distance between two adjacent bubbles, in this case of relatively equal size, falls below a certain critical level, the film rupture leads to the coalescence of the bubbles, resulting in a new round bubble in a time frame of 1-2 ms, as shown in Fig. \ref{fig:merging}. 
The volume of this new bubble is obviously equal to the sum of the volumes of the previous bubbles. 
As the bubbles involved are relatively small compared to others and to the foam size, no significant bubbles or foam rearrangements are observed in the surrounding area in this case. 
Similar behavior has been observed in the past with other alloys with X-ray radioscopy \cite{met2010010}. 
The kinetics of coalescence are random in nature but influenced by various parameters such as gravity or temperature \cite{SoftMatter:2011}. 
To gain a more detailed insight into this coalescence process, even faster acquisition speeds will be required. 
This conclusion is consistent with the X-ray tomoscopy results for a more stable liquid metal foam alloy \cite{García-Moreno2019}.

\begin{figure*}[htbp]
\includegraphics[width=1\textwidth]{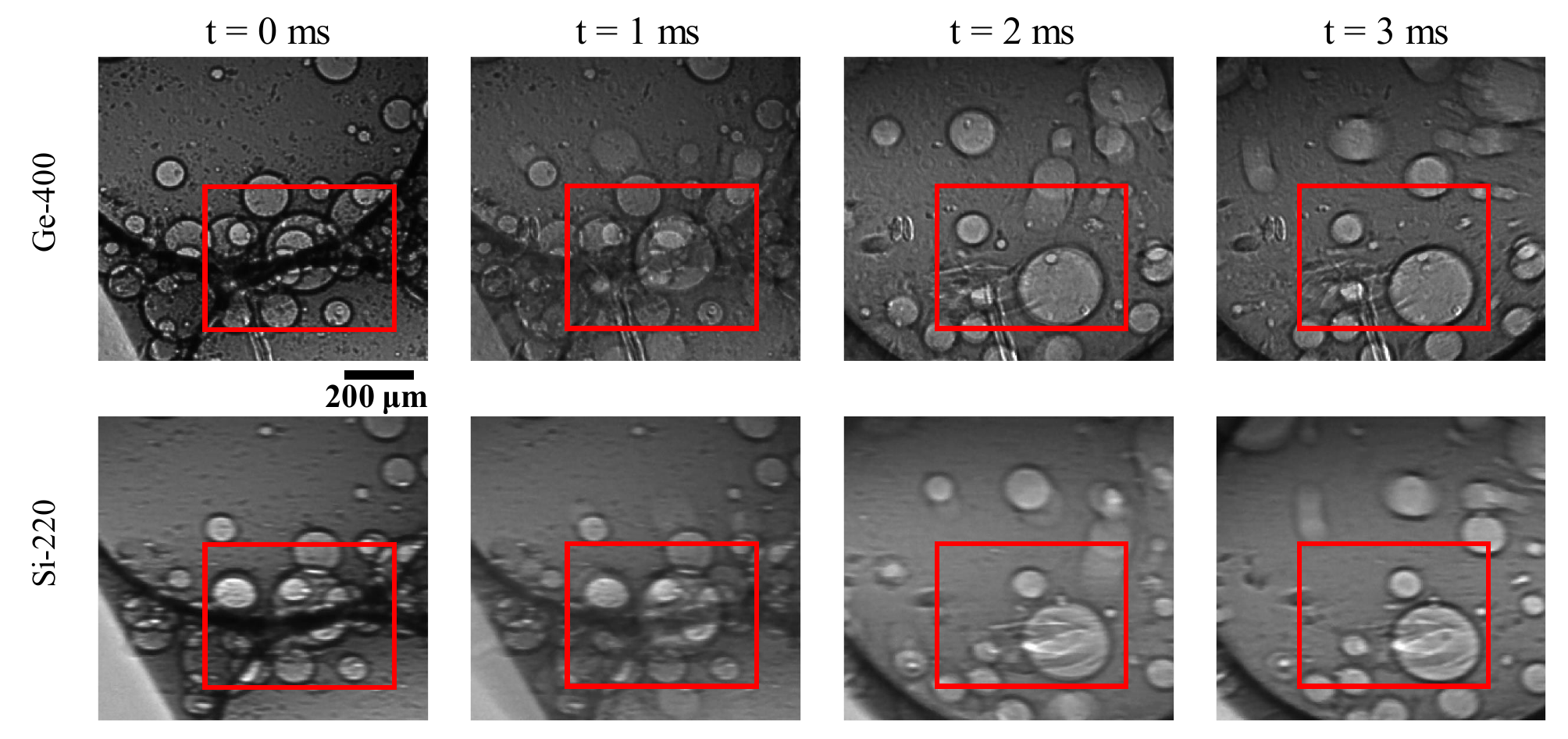}
\caption{
Selection of frames showing bubble coalescence and rupture in the melted aluminum foam sample. 
The red square marks the region of interest where a film between two relatively large bubbles ruptures, leading to coalescence of the two bubbles involved, and also to an avalanche of further bubble coalescence and structural rearrangements. 
 }  
\label{fig:rupture}
\end{figure*}

If the bubbles involved in a coalescence process are large relative to the foam size, as shown in Fig. \ref{fig:rupture}, the film rupture process leading to coalescence of the bubbles can be accompanied by a coalescence avalanche \cite{met7080298} and a vigorous structural rearrangement of the entire foam, which directly affects the development of foam expansion. 
The film rupture is again faster than 1 ms and the coalescence of the two bubbles takes place in a few milliseconds.

\begin{figure*}[htbp]
\centering
\includegraphics[width=.8\textwidth]{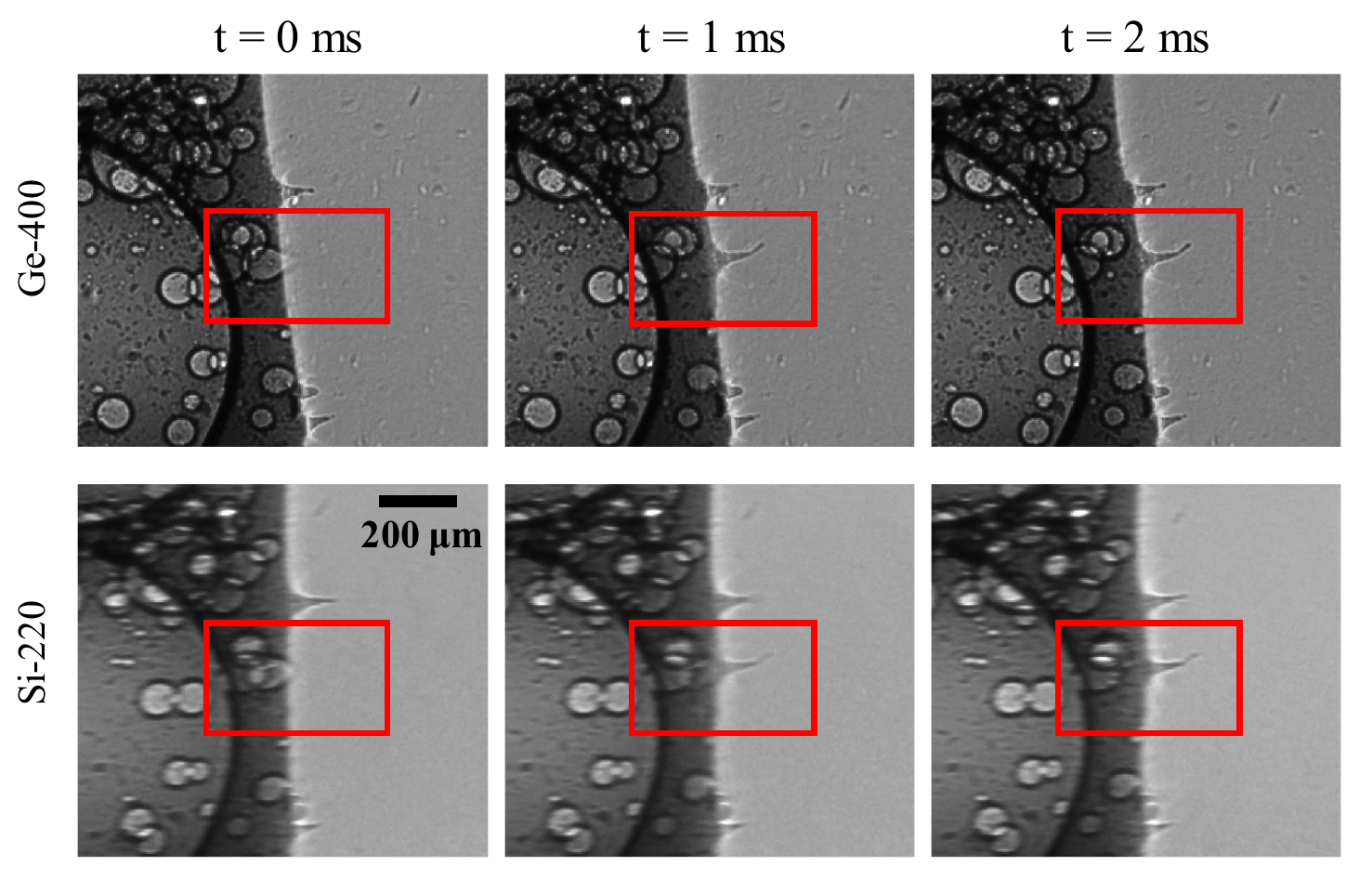}\caption{
A selection of magnified frames showing the rupture of a film at the foam surface leading to the burst of a bubble and the generation of a surface spike in the course of fast solidification. 
 }  
\label{fig:spike}
\end{figure*}

A new phenomenon, namely the formation of spikes on the foam surface, was observed and is shown in a sequence in Fig. \ref{fig:spike}. 
When a bubble bursts near the surface, a film rupture occurs at the outer foam skin, with molten material spurting out in less than a millisecond. 
The liquid appears to be held in a kind of spike-like surface protrusion by a solid oxide skin and solidifies in 1-2 ms. 
The formation of such spikes could only be observed in cases where foam growth and subsequent foam solidification are extremely rapid, as is the case here.

\section{Conclusions}
This experiment constitutes the first implementation of XMPI at a diffraction-limited storage ring using pink beam, i.e., the full-harmonic provided by an insertion device. 
The XMPI technique, coupled with the high spectral density and full-harmonic beam of the ID19 beamline, allowed to successfully image sub-millisecond dynamics (3000 frames per second) with spatial resolution of $8$ µm per projection while filling the full dynamic range of the detectors.
The narrow energy bandwidth of this implementation enables the images to carry quantitative information on the sample, as well as produce sharper images compared to a white beam.
Thus, the results here presented are a fundamental step to establish XPMI as a technique capable of probing 4D (3D+time) phenomena in samples opaque to visible light with kHz frame-rate speed and micrometer resolution that are sensitive to shear forces. 

XMPI was employed for the study of the dynamics of the aluminum foaming process at 1~kHz, avoiding shear forces that alter the studied dynamics as introduced by time-resolved tomography in such systems.
From the acquired dataset, 3D movies of bubbles moving in the melted aluminum were obtained via machine learning algorithms. 
The absence of shear forces was crucial to enable the observation of a novel phenomenon in aluminum foaming, namely the formation of spikes on the foam surface. 
These spikes are formed by the bursting of small bubbles near the sample surface and solidify quickly due to the colder ambient temperature and the formation of an oxide layer. 
Furthermore, the rupture of small and large films separating two bubbles and leading to their coalescence could be investigated for the first time at 1~kHz in a liquid AlSi6Cu4 alloy foam. 
In agreement with other results in the literature, film rupture times of less than 1 ms could be detected, indicating the need for even faster acquisition speed. 

As a result of this implementation, we are designing a beam-splitting scheme tailored to the beam characteristics of ID19 that will allow additional projections with increased angular spacing between each projection to benefit the 3D reconstruction process. 
Furthermore, this future implementation of the XMPI instrument will increase the spatiotemporal resolution by enhancing the efficiency and image quality of the crystal splitters by using symmetric Bragg diffraction and improving the mounting and crystal manufacture. 



\section{Backmatter}
\begin{backmatter}

\bmsection{Funding} 
This work was performed within the following projects: RÅC (Röntgen-Ångström cluster) “INVISION” project, 2019-2023; EuXFEL R\&D “MHz microscopy at EuXFEL: From demonstration to method”, 2020 - 2022; 
ERC-2020-STG, 3DX-FLASH, Grant agreement 948426;
Horizon Europe EIC Pathfinder “MHz-Tomoscopy” project, 2022-2025.
The Czech co-authors‘ thank the Czech Ministry of Education, Youth, and Sports (CMEYS) for partial financial support of the study and utilization of measurements and control of radiation-induced thermomechanical phenomena at large-scale facilities, as a part of the grant nr. LM2023068. 
The Slovak co-authors‘ thank the grant titled High-performance curved X-ray optics prepared by advanced nanomachining technology from Scientific Grant Agency of the Ministry of Education, science, research and sport of the Slovak Republic and the Slovak Academy of Sciences,  no.: VEGA - 2/0041/21.

\bmsection{Acknowledgments} 
We thank the team of ESRF ID19 for the help during beamtime MI1449.

\bmsection{Disclosures} The authors declare no conflicts of interest.

\bmsection{Data availability} Data underlying the results presented in this paper are not publicly available at this time but may be obtained from the authors upon reasonable request.

\bigskip


\end{backmatter}

\bigskip

\bibliography{References}

\end{document}